**Post-Impact Thermal Evolution of Porous Planetesimals**

Thomas M. Davison[1*], Fred J. Ciesla[1] and Gareth S. Collins[2]

[1]Department of the Geophysical Sciences, University of Chicago, 5734 S Ellis Avenue, Chicago, IL 60637, USA

[2]Department of Earth Science and Engineering, Imperial College London, London, SW7 2AZ, UK

[*]Corresponding author email address: tdavison@uchicago.edu (T.M. Davison)

# ABSTRACT

Impacts between planetesimals have largely been ruled out as a heat source in the early Solar System, by calculations that show them to be an inefficient heat source and unlikely to cause global heating. However, the long-term, localized thermal effects of impacts on planetesimals have never been fully quantified. Here, we simulate a range of impact scenarios between planetesimals to determine the post-impact thermal histories of the parent bodies, and hence the importance of impact heating in the thermal evolution of planetesimals. We find on a local scale that heating material to petrologic type 6 is achievable for a range of impact velocities and initial porosities, and impact melting is possible in porous material at a velocity of > 4 km/s. Burial of heated impactor material beneath the impact crater is common, insulating that material and allowing the parent body to retain the heat for extended periods (~ millions of years). Cooling rates at 773 K are typically 1 - 1000 K/Ma, matching a wide range of measurements of metallographic cooling rates from chondritic materials. While the heating presented here is localized to the impact site, multiple impacts over the lifetime of a parent body are likely to have occurred. Moreover, as most meteorite samples are on the centimeter to meter scale, the localized effects of impact heating cannot be ignored.

# 1. INTRODUCTION

Impacts between asteroid-sized bodies have been common occurrences throughout the history of the Solar System and have played a role in shaping the chemical and mineralogical properties of these objects throughout their lifetimes. Impact melt in meteoritic materials is widespread (e.g. Bischoff et al. 2006; Weirich et al. 2010) demonstrating that high temperatures can be reached in such events. These melts are thought to result from collisions into the meteorite parent body, at velocities of kilometers per second. In addition to melting, shock metamorphism has been recognized in meteorites (Stöffler et al. 1991; Sharp and de Carli 2006), again providing evidence of the collateral effects that collisions have on the chemistry and mineralogy of minerals in a given parent body.

The importance of such collisions in the early evolution of the solar system is debated. On one hand, impacts that led to the impact melt seen in meteorites would have been much more common and energetic during the epoch of planet formation than they are today. Not only was the number of small bodies much greater in the first 100 Ma of our Solar System's formation, but also gravitational stirring by growing and fully formed planets and their precursors would have dynamically excited planetesimals. For example, in the terrestrial planet region, impact velocities would range from 1-10 km/s (Bottke et al. 2005). This has led some to postulate that such collisions could be responsible for at least some of the early metamorphism and melting seen in meteoritic materials (Wasson et al. 1987; Rubin 1995; Keil et al. 1997; Rubin et al. 2001).

On the other hand, a number of workers have shown that much of the thermal metamorphism and melting in primitive bodies can be explained solely by the decay of short-lived radionuclides, such as $^{26}$Al (McSween et al. 2002; Hevey and Sanders 2006; Grimm and McSween 1989; Trieloff et al. 2003; Ghosh et al. 2006; Kleine et al. 2008; Harrison and Grimm 2010). Specifically, these models have tracked the rate of heat produced by the decay of the short-lived radionuclides, its redistribution in the parent body, and its loss to space at the surface. The peak temperatures and cooling rates reached by different portions of the modeled parent bodies are consistent with those inferred from the degree of metamorphism and metallographic cooling rates recorded by different meteoritic samples, although Harrison and Grimm (2010) were unable to match data for some chondrites with radiogenic heating alone.

To date, no similar study has been carried out to investigate the type of thermal evolution expected as a result of impacts. The most relevant study to date is that of Keil et al. (1997). These authors examined terrestrial impact craters and carried out numerical simulations of the collision of asteroid-sized bodies to evaluate the importance of heating in such events. As the amount of melt and metamorphosed rock around a terrestrial impact crater is small compared to the volume of the crater, and because numerical models showed that impacts would not cause a globally averaged temperature increase in a planetesimal of more than 10 - 100 K, Keil et al. (1997) concluded that impacts were not important sources of the heat needed to explain the metamorphic and igneous processes recorded by meteorites.

While Keil et al. (1997) ruled out impacts as being an important source of heating on the global scale, impacts may still be important in producing local thermal effects. Recent studies have quantified some of the thermal effects of large-scale collisions (Asphaug et al. 2006; 2010), such as igneous alteration, degassing, petrogenesis and (possible) melting that occurred in so called 'hit-and-run' collisions between differentiated Moon to Mars sized planetary bodies. Davison et al. (2010) quantified the amount of heating and melting produced in collisions at velocities of kilometers per second between planetesimals of the type expected in the early solar system and showed that localized heating is possible. However, little work has been done to investigate the long-term thermal effects of collisions between planetesimals (i.e. for millions of years after the collision).

The extent of the thermal anomalies produced in collisions of the type studied in Davison et al. (2010) depends on the porosity of the bodies involved, as the energy of a collision can be locally concentrated by the work required to close pore space (Zel'Dovich and Raizer, 1967). While Keil et al. (1997) discussed porosity they only considered its effects on the disruption threshold of the parent body, and not on its shock physics implications. Such effects must be accounted for, as planetesimals were likely highly porous after their formation. While it is unclear exactly how planetesimals formed in the early solar system, it must have been a gentle process such that the dust grains or aggregates from which they formed collided at relatively low velocities, either by continuous aggregation (Weidenschilling 1980; 1984; 1997), gravitational contraction in a clump due to the streaming instability (Johansen et al. 2007; 2009) or turbulent concentration (Cuzzi et al. 2001; 2008; 2010). In the case of aggregational growth, collisions must have occurred at low enough velocities to allow net growth. In the case of gravitational clumping, the relative velocities of colliding particles must be small enough for them to remain gravitationally bound. At the collisional velocities expected, Teiser and Wurm (2009) have shown that aggregates would be highly porous (> 60%), and Cuzzi et al. (2008) even referred to the planetesimals that would form from gravitational clumps as "sandpiles". More recently, Bland et al. (2011) used the observed fabric of minerals in the matrix of a chondritic meteorite to infer the porosity of the primitive matrix material. They noted a preferred orientation of grains in the matrix, most easily explained by reorientation of a random grain distribution in a single, unidirectional compaction event. The observed degree of grain rotation was used to quantify the amount of volume lost during compaction (~50%) and conclude that the precursor material must have had a much higher initial matrix porosity (70 - 80%), relative to the ~40% pore space now present in the meteorite matrix.

Porosity can have significant effects on the outcome of a collision, as much of the impact energy is consumed in the process of collapsing pore space. First, the presence of porosity can reduce the size of the crater formed in a collision when compared to an impact into a non-porous target (e.g. Love et al., 1993; Wünnemann et al., 2006). Second, the extra energy required to close pore space means that the shock wave created in the collision is attenuated more rapidly, and this a smaller volume will be subject to high shock pressure. However, a third competing effect of porosity is that, to produce the same post shock temperature, a lower shock pressure is required in a porous material compared to a non-porous material. The study of Davison et al. (2010) investigated the interplay of the last two of these effects, to determine the influence of porosity on heating in collisions between planetesimals, and showed that while heating on the global scale was unlikely from all but the most energetic of collisions, on the local scale (i.e. ~ the scale of the impact crater), the presence of porosity could greatly increase the peak temperatures and mass of material heated by a collision. That study also showed that the impact velocity required to heat material to a given temperature is lower for porous material: for example, in a collision between two equal-sized, non-porous planetesimals, a velocity of ~ 12 km/s was required to heat the total mass to the solidus, whereas in bodies of 50% porosity, this was achieved at ~ 7 km/s.

Here we perform numerical simulations to quantify, in detail, the global and local processing that would have occurred as a result of an impact on a young planetesimal during the early evolution of our solar system. The iSALE shock physics code allows us to account for changes in porosity and temperature of the colliding materials during the collision event. We then take the resulting physical and thermal structure of the surviving planetesimal and track the redistribution of the energy within the planetesimal and its loss to space at the surface, as is done in the thermal studies of short-lived radionuclides. This provides quantitative information concerning the extent and level of thermal evolution that would be caused in a planetesimal as a result of impact. We then compare the predictions of our models to data reported in the literature on the thermal evolution of meteoritic materials.

In the next section we describe the hydrocode simulations we performed as part of this study. We then present the subsequent thermal evolution of the surviving bodies in these collisions. We end with a discussion of our results and the implications for the role of impacts in the thermal evolution of planetesimals in the early solar system.

## 2. IMPACT MODELING

### 2.1. Methods

In this study, we model the collision between two planetesimals using the iSALE shock physics code (Wünnemann et al. 2006), which is an extension of the SALE hydrocode (Amsden et al. 1980). To simulate impact processes in solid materials several modifications were made to the original SALE code, including: the addition of an elasto-plastic constitutive model, fragmentation models and various equations of state (EoS), and the ability of the code to model multiple materials (Melosh et al. 1992; Ivanov et al. 1997). Development of the code is ongoing: recent improvements include a modified strength model (Collins et al. 2004) and a porosity compaction model (Wünnemann et al. 2006; Collins et al. 2011a). iSALE has been extensively validated against laboratory experiments and other hydrocodes (e.g. Pierazzo et al. 2008). It has previously been used to simulate collisions between porous planetesimals (Davison et al. 2010).

The simulations in this work differ slightly from those used in Davison et al. (2010). In that work, the aim of the simulations was to determine the mass of material shock heated to several post-shock temperatures. Lagrangian tracer particles were used to track the pressure history of the material. To determine the mass of material shock heated to a certain post-shock temperature, the peak-shock pressures of the tracer particles could be examined early in the simulation – once the shock wave had been fully released. By summing the masses of the tracer particles shocked to a given critical pressure, the mass of material that would reach a given post-shock temperature could be easily determined. It was therefore not necessary to run the models past the cycle of shock and release. In this work, we are concerned with the longer-term thermal evolution of the heated material, and hence we must simulate a longer time period after the collision. This allows us to determine the final position of the heated material and quantify the compaction throughout the parent body and the ejection of heated material; to achieve this, it is necessary to simulate the self-gravity of the colliding planetesimals.

The gravity field in the simulations presented in this work was updated periodically during the calculation using a self-gravity algorithm (Collins et al. 2011b), based on that described by Barnes and Hut (1986). During a gravity field update the gravitational attraction of computational cells containing mass (actually rings of mass due to the axial symmetry employed in the two-dimensional iSALE model) is calculated for each vertex of the mesh. Computational efficiency is achieved with minimal loss of accuracy by combining cells into patches of 2x2, 4x4, 8x8, etc. neighboring cells when they are sufficiently far from the point of interest, and computing the gravitational acceleration of the patch (ring) as a whole. The size of the patch $h$ is determined by a user-defined accuracy parameter, the opening angle $\theta$, times the distance $l$ between the patch center and the vertex at which gravity is being computed, $h = \theta l$ (Barnes and Hut 1986). Thus, large numbers of cells are combined to form massive patches (rings) when accounting for the gravitational acceleration of mass far from the point of interest. The self-gravity algorithm was tested against benchmark problems discussed in Crawford (2010). The gravity field was updated using this method once every 50 computational cycles in the simulations discussed in this work. The timestep for each computational cycle is adaptive in iSALE, but typically this corresponded to a gravity field update approximately every 2 seconds.

*2.1.1. Material model*

An analog material was used to represent the planetesimal material: Dunite, the olivine-rich endmember of peridotite, has previously been used to represent chondritic material (Davison et al. 2010), and is a reasonable approximation for the chemical composition of planetesimals. Dunite also has the benefit of having a well-defined set of input parameters for the ANEOS equation of state package (Benz et al. 1989). An ANEOS equation of state table for dunite was therefore used to represent the thermodynamic response of the non-porous component of the planetesimal material. Material was assigned a shear strength using the procedure outlined in Collins et al. (2004), with strength parameters for weak rock (e.g. Leinhart and Stewart, 2009). Pore-space compaction was simulated with the $\varepsilon$-$\alpha$ porous-compaction model (Wünnemann et al., 2006) including the recent improvements to include the thermal expansion of the matrix during compaction (Collins et al., 2011a). Based on previous studies (e.g. Wünnemann et al. 2008; Davison et al. 2010), the porous-compaction parameter, $\kappa$, was set to 0.98.

One limitation of the continuum approximation used in the porous-compaction model is that it requires the scale of the pores to be smaller than the scale of the computational cells, and for the pores to be uniformly distributed throughout the material. Therefore, the model assumes that any heating and compaction are averaged over the bulk material. In natural materials, pore space is often heterogeneously distributed; thus heating by shockwaves can lead to localized "hotspots" (Kieffer et al., 1976), on the scale of the pores, which cannot be resolved by this model.

*2.1.2. Initial and boundary conditions*

iSALE was used in its two-dimensional, axi-symmetric mode. Two collision geometries were studied: First, a collision with an impactor-to-parent body radius ratio ($R_i/R_{pb}$) of 0.1 ($R_{pb}$ = 250 km; $R_i$ = 25 km), for which the self-gravity algorithm was employed. Second, a set of simulations were performed with $R_i/R_{pb}$ = 0.01. In this suite of simulations, only a portion of the parent body proximal to the impact event was modeled (to a depth and radial distance of 150 km), as the shock wave was found not to influence the parent body material beyond that distance from the impact site. As the impact had little effect on the bulk properties of the parent body, the gravity field for these simulations was calculated once at the start of the calculation and not updated. All material was assigned an initial temperature of 300 K, and a constant porosity value ($\phi$ = 0, 20, 50%). The parent body material was assigned no initial velocity, and all cells within the impactor were assigned a velocity normal to the parent body surface at the point of impact ($v_i$ = 4 km/s, 6 km/s). These velocities are comparable to the current relative velocities of the asteroid belt (Bottke et al. 1994), and would easily have been achieved during the early stages of the Solar System (for example, during planet formation, small bodies were scattered to eccentric orbits; as $V_{imp} \sim eV_{orbit}$, eccentricities of 0.2 - 0.3 in the asteroid belt would produce such collisional velocities). Only normal incidence collisions were considered in this work, so that the simulations could be run in axial symmetry (to save on computational expense). An investigation into the effect of impact obliquity is in progress; however, Pierazzo and Melosh (2000) showed that melt volume should scale according to the volume of the transient crater. For an impact at 45° (the most common impact angle), the heated mass may be reduced by ~35% from the figures presented here. The boundary condition on the symmetry axis was a free slip condition (i.e. on the boundary there is no normal flow, but material is allowed to move along the boundary), and all other domain boundaries were assigned outflow conditions, where material is allowed to flow freely across the boundary (at which point it is deleted). Impacts would have occurred that spanned a wider range of parameters than those considered here. We focus on this subset in order to document the effects of collisions into porous planetesimals and save a larger parameter exploration for future work.

*2.1.3 Resolution*

In simulations with $R_p/R_{pb} = 0.1$, the parent body was modeled with a resolution of 200 computational cells per body radius (i.e. for a 250 km radius parent body, the resolution was ~ 1 km per cell); the resolution of the impactor in the $R_p/R_{pb} = 0.1$ simulation was 20 cells per impactor radius. In the simulations with $R_p/R_{pb} = 0.01$, the impactor was modeled with a resolution of 20 cells per impactor radius, and the parent body had 2000 cells across its radius.

## 2.2. Results

Figure 1 shows the formation of the impact crater in two simulations. Figure 1(a) shows the initial conditions of an impact between a 25 km radius impactor with a 250 km radius parent body. The initial porosity is $\phi = 0.5$, and the impact velocity is 4 km/s. Figure 1(b) shows the opening of a transient crater 500 s after the impact. After 1500 s (Figure 1(c)) the crater starts to collapse due to the gravity of the parent body; the strongly heated impactor material that was lining the crater floor starts to form a plug in the center of the crater floor. A region of compacted, heated parent body material exists beneath the crater floor. Figure 1(d) shows a point in time later in the collapse process, during which time the heated impactor material has formed a hot (~ 1200 – 1400 K) plug of material. After 10000 s (Figure 1(e)), the collapse of the crater is complete, and the heated region has been buried by cooler, less compacted material from the parent body; the heated plug is buried by ~ 12 km of cooler material. The heated, compacted plug of impactor material extends to a radial distance, $r$, of ~ 20 km, comparable to the radius of the impactor. The parent body material does not reach such high temperatures, but a heated, compacted region extends over a larger radial distance: material at a radial distance of 30 km is heated to 800 K, at $r = 40$ km to 700 K, and at $r = 65$ km to 400 K. For a more detailed view of the heated region, see Figure 3(a).

Figure 1(f) shows the initial condition for an impact of a 2.5 km radius impactor into the same sized parent body (250 km radius). Figure 1(g) shows the opening of a transient crater. In Figure 1(h), the crater has stopped growing, and the heated impactor material has collected at the center of the crater floor. In this case, a simple impact crater has formed where less gravitational collapse takes place. The modification stage of the impact event is limited to slumping on the walls of the crater (Figure 1(i)). The heated material is buried by this slumped material in Figure 1(j) to a depth of ~ 1 km. The heated plug of impactor material has a radius of 2.3 km. The radial extent of the heated parent body material is: 800 K at $r = 3.5$ km, 700 K at $r = 4.5$ km, and 400 K at $r = 7.0$ km.

Final craters of other impact events with velocities of 4 km/s and 6 km/s and porosities of $\phi = 0.0$, 0.2 and 0.5 are shown in Figure 2 for the two impact geometries discussed here. In general, the heated regions are larger and the peak temperatures are higher for the faster impact velocity (Figure 2(c–e, h–j)). The morphology of the heated region is similar for $\phi = 0.2$ (Figure 2(a, d, f, i)) and $\phi = 0.5$ (Figure 1(e, j) and Figure 2(c, h)).

For impacts with non-porous materials, the peak temperatures are lower and the heated regions are smaller. Also, and critically, less burial of the heated material occurs in the non-porous collisions. For example, in the collision of a 25 km radius impactor at 4 km/s, shown in Figure 2(b), the heated plug is smaller, as less of the heated impactor is retained on the surface. The depth to the bottom of the heated plug is ~ 25 km, and at its widest point, it has a radius of ~ 40 km. Some impactor material at top of the plug appears exposed to the surface: i.e., the burial depth is at the limit of the resolution of the model. The temperatures of the plug of impactor material is ~ 1000 – 1100 K. In the parent body, the radial extent of the heated region is ~ 80 km for material heated to 400 K, ~ 65 km for material heated to 500 K, and ~ 45 km for material heated to 600 K.

# 3. POST-IMPACT COOLING

## 3.1. Methods

Once the planetesimal had relaxed to its final state, we noted the temperature and density in each cell and imported these values into a separate thermal evolution code to track how the energy was redistributed with time. The code follows the approach used in thermal models that have been used to investigate the effects of short-lived radionuclides, as it solves the heat diffusion equation throughout the surviving planetesimal, with the surface temperature of the planetesimal set by the ambient nebular temperature (170 K). However, there are two major differences between our approach and those adopted in the previous studies. The first is that we do not allow for any heat input from short-lived radionuclides. This is not meant to suggest that such effects are unimportant. Rather the goal here is to quantify the thermal effects due to collisional processes alone. Once the thermal evolution caused by impacts is understood, the combined effects of impacts and short-lived radionuclides can be studied. We leave these combined effects as the subject of future work.

The second difference is that because the parent body is no longer spherically symmetric after the impact, we solve the heat diffusion equation in two dimensions using cylindrical coordinates:

$$\frac{\partial T}{\partial t} = \frac{1}{r}\frac{\partial}{\partial r}\left(\kappa r \frac{\partial T}{\partial r}\right) + \frac{\partial}{\partial z}\left(\kappa \frac{\partial T}{\partial z}\right)$$

where $\kappa$ is the thermal diffusivity of the materials. The thermal diffusivity is related to other material properties through

$$\kappa = \frac{K}{\rho C_p}$$

where $K$ is the thermal conductivity of the materials of interest, $\rho$ is the density of the material, and $C_p$ is the heat capacity. For each planetesimal, the thermal model uses the same grid construction, and thus spacial resolution, as used in the iSALE simulations.

For the thermal properties, we assume a constant heat capacity of $8 \times 10^6$ erg/g. While in reality the heat capacity is a function of temperature and specific make-up of the planetesimal, this value serves as a representative value for solar system silicates and gives a similar thermal evolution to that produced by a temperature dependent expression (Ghosh and McSween, 1999). The density in each grid cell is taken from the iSALE simulations. The thermal diffusivity is taken to be $1 \times 10^{-3}$ cm$^2$/s, the same value used by Ghosh et al. (2003). Porous materials are more insulating than solid ones, as thermal conduction can occur only between materials that are in contact. For porous materials, we adopt the scaling based on experiments on lunar rocks and soils of Warren (2011), which argues that the thermal conductivity varies with porosity, $\phi$, as:

$$K(\phi) = K_0 e^{-12.46\phi}$$

This relation gives a thermal diffusivity that then scales as:

$$\kappa(\phi) = \kappa_0 \frac{e^{-12.46\phi}}{(1-\phi)}$$

as the density also decreases by a factor of $(1 - \phi)$, with $\phi$ being the fractional porosity of the planetesimal ($\phi = 1 - \rho_{porous}/\rho_{non-porous}$). In these expressions $0 \leq \phi \leq 1$. Planetesimal porosity had been considered in previous thermal models, whether it be via an insulating layer of porous regolith on the surface of a body as was done by Ghosh et al. (2003) or the original structure of the entire body as discussed by Sahijpal et al. (2007). In these cases, the porous material (no value of porosity was given) was assumed to have thermal diffusivities that were 1/100$^{th}$ and 1/1000$^{th}$ of those of the non-porous material respectively. This would correspond to porosities of ~ 0.4 and ~ 0.65 respectively with the functional forms used here, which is at the upper end of the porosities we consider in our work (we consider a maximum of 0.5).

To test our model, we calculated the thermal evolution of a perfectly spherical planetesimal that was heated by the decay of $^{26}$Al. The thermal evolution of such a body, assuming constant thermal properties, has an analytic solution (Carlslaw and Jäger, 1959) to which our numerical model could be compared. We found very good agreement between our model and the analytic solution—for example, in the case of a 100 km diameter body formed 2 million years after CAIs (thus defining the amount of live $^{26}$Al present) and broken up into cells of 1-km side length in both the $r$ and $z$ directions, the temperatures in our model were within 2% of those predicted by the analytic solution. There were no significant differences in the thermal profiles for the $r$ and $z$ directions. We compared a number of cases against the analytic solution and found similar agreement with the analytic solution. We are confident that our thermal model provides an accurate description of how energy is redistributed within the planetesimal.

## 3.2. Results: Thermal history

An example of the detailed thermal evolution of one of the impact simulations is shown in Figure 3, corresponding to the impact scenario shown in Figure 1(a-e; $R_i$ = 25 km, $v_i$ = 4 km/s, $\phi$ = 0.5). Figure 3(a) shows the temperature field immediately after the impact (equivalent to $t$ = 10000 s in Figure 1(e)). ~ 10 km of weakly heated, uncompacted parent body material insulates the more strongly heated and compacted impactor material at this time. After 2 Ma (Figure 3(b)), the hot plug of impactor material remains at > 1350 K (remaining at this temperature for such a prolonged period may lead to some coagulation of molten metal and sulfide). The hot material that is buried loses heat by conduction. The highly porous (uncompacted) material surrounding the compacted and heated zone is a poor conductor and acts as an effective buffer to heat transport. As a result, heat losses from the hot region occur predominantly via conduction toward the surface, through the (more thermally conductive) compacted zone. The hottest region for much of the simulation is the base of the compacted zone, which is surrounded by porous material unaffected by the impact. After 10 Ma, the peak temperature is ~ 1250 K (Figure 3(c)), after 50 Ma, the peak temperature at the center of the heated region is 950 K, and after 100 Ma, the peak temperature is ~ 820 K. An approximately spherical region of radius 20 km is heated above 780 K, and material at a radial distance of 40 km remains heated above 650 K. Material at $r$ = 65 km remains heated to 400 K, or 100 K above the initial temperature, at these late times.

Figure 4 shows the thermal evolution of the simulation presented in Figure 1(f-j; $R_i$ = 2.5 km, $v_i$ = 4 km/s, $\phi$ = 0.5). Only approximately 1 km of parent body material covers the strongly heated impactor material at $t$ = 0 Ma, which means that the heated region cools much more rapidly than in the case with $R_i$ = 25 km discussed above. Heat is retained for a short period of time: after 0.2 Ma, the peak temperature has decreased from 1350 K to 1200 K; however, by 1 Ma after the impact, the peak temperature is 900 K and by 4 Ma the peak temperature is 600 K.

### 3.3. Results: Peak temperatures and cooling rates

Figures 3 and 4 demonstrate that the heat deposited by an impact can be retained for extended periods of time, from a few million years to > 100 Ma. These are the same timescales discussed when considering radiogenic heating models. Whether impacts can provide the type of thermal evolution needed to explain the observed features of meteoritic materials requires closer examination.

Several types of data are available from the meteoritic record to describe the thermal evolution of a parent body. In this work, we compare results from our thermal models with two of those measurements: the peak temperatures reached by the meteoritic material and the cooling rate of that heated material. The peak temperature is used to define the petrologic type of heated meteoritic material. In this work, we use the definitions of Harrison & Grimm (2010) for type 3 ($T_{peak}$ < 948 K), type 4 - 5 (948 K < $T_{peak}$ < 1138 K) and type 6 (1138 K < $T_{peak}$ < 1273 K). We define type 7 as mass heated above 1273 K, but below the dunite solidus (1373 K; McKenzie and Bickle 1988), and we use 2053 K as the dunite liquidus (Katz et al. 2003). For all the peak temperatures presented in this work, the material remained at that temperature for at least one year. The cooling rate is defined as the rate at which the material was cooled as it passed through a particular temperature during its thermal evolution. Many data have been published on cooling rates of meteorite samples. One such measurement, the metallographic cooling rate (Wood, 1967), is used to infer the cooling rate at 500°C (773 K). Here we summarize some key findings from the literature of typical metallographic cooling rates.

Taylor et al. (1987) published a large number of cooling rate measurements for H, L and LL meteorites. Cooling rates ranged from 1 K/Ma to > 1000 K/Ma. Taylor et al. (1987) found no correlation between petrologic type and cooling rate, for each of the three parent bodies that they studied. Harrison and Grimm (2010) summarize cooling rate measurements for the H-chondrite parent body (from Taylor et al. 1987; Willis and Goldstein 1981; Williams et al. 2000), in which they show a range of cooling rates from 2 – 50 K/Ma for type 3 H-chondrites, from 4 – > 1000 K/Ma for type 4 – 5 H-chondrites, and 5 – 75 K/Ma for type 6 H-chondrites, although they excluded various measurements, especially high cooling rates. With this inclusion of those cooling rates, the range of cooling rates may have been even greater. Also, metallographic cooling rate measurements have an uncertainty of a factor of 2.5 (Taylor et al. 1987). Thus, a thermal model that produces cooling rates in the range of 1 – 1000 K/Ma would be consistent with the measurements from the variety of metallographic cooling rate studies listed above. Recent studies of the onion shell structure (e.g. Harrison and Grimm 2010) have attempted to fit thermal models to cooling rate measurements and found in models that best fit observational constraints cooling rates (at 773 K) of ~ 0 – 50 K/Ma. Kleine et al.

(2008) modeled $^{26}$Al decay to fit measurements from the Hf-W thermal chronometer in H-chondrites, and derived cooling rates from their model. Cooling rates at 773 K were not presented in that work, but at 723 K the cooling rates ranged from ~ 5 – 55 K/Ma. One goal of this work is to determine the range of cooling rates possible in a parent body after being heated in a collision, and to compare this to the observations summarized above.

In performing our post-impact thermal evolution, we recorded the peak temperatures reached by each cell of the surviving planetesimal, the mass in each cell, and the cooling rate at 773 K (the temperature recorded by metallographic cooling rates). These data are listed in Tables 1 and 2 and shown in Figures 5 – 9. Table 1 shows, for each simulation, the cooling rate of material from the impactor and parent body (mean and standard deviaton). Table 2 shows, for each simulation, the mass of material heated to each petrologic type, split into the mass from the impactor, and the mass from the original parent body. Heated masses quoted in this section, and in Table 2, are normalized by the impactor mass, $M_i$, and are for a parent body of radius 250 km. Only material that was retained on the parent body after impact was considered in this analysis. In all cases presented below, less than 0.01% of the parent body mass was ejected from the surface. As heat was conducted to the surface, some material reached a peak temperature higher than its immediate post-impact temperature. Figure 5 shows this increase in temperature for the three simulations with $v_i = 4$ km/s and $R_i = 25$ km. In the non-porous case, this effect seems to be small: a few grid cells are heated by ~ 200 K after the impact, but the majority of cells see little (< 50 K) or no post-impact heating, as heat is quickly lost to space. In the $\phi = 0.5$ case, some material is heated by > 600 K after the impact, and a large number of computational cells are heated by > 100 K, as heat is diffused through the body.

For comparison with other models of planetesimal thermal evolution, the mean and standard deviation of the cooling rates for all cells are calculated, to illustrate the typical range of cooling rates predicted by the model. The cooling rates are approximately log-normally distributed, and hence the geometric mean and standard deviation are used. In some cases, the standard deviations presented here are large. It should be noted that these do not represent large errors in the modeling, but rather show that it is possible for a single impact event to produce a wide range of cooling rates in meteorites of a given petrologic type.

*3.3.1. $R_i = 25$ km and $v_i = 4$ km/s*

Figure 6(f) shows the peak temperatures for the $\phi = 0.5$ simulation of a 25 km impactor into a 250 km body (see Figure 1(a-e) and Figure 3). In this simulation, most of the impactor material (~ 0.85 of the impactor mass, $M_i$) is heated to at least petrologic type 6, with some material (~ 0.12 $M_i$) heated above the dunite solidus (labeled 'solid + melt' on Figure 6). The parent body is not as strongly heated: 0.36 $M_i$ of the parent body material is heated to type 4 - 5, and 0.02 $M_i$ is heated to type 6 or above. The mean cooling rate of the impactor is $1.16^{+0.27}_{-0.22}$ K/Ma, and for the parent body is $1.00^{+0.46}_{-0.32}$ K/Ma (Figure 6(c)). Approximately 0.4 $M_i$ of parent body material is ejected during the collision, corresponding to $4\times10^{-6}$ $M_{pb}$.

For the same collision geometry with $\phi = 0.2$, no material is heated above the solidus, from either the impactor or the parent body. This is expected, as lower porosity leads to lower temperatures in these collisions (Davison et al. 2010). Of the impactor, ~ 0.09 $M_i$ is heated to type 4 – 5, 0.76 $M_i$ is heated to type 6, and ~ 0.01 $M_i$ is heated to type 7. 0.07 $M_i$ is heated to type 3, and the remaining ~ 0.07 $M_i$ is ejected from the parent body during the impact event. Very little of the parent body (< 0.01 $M_i$) is heated to type 6. ~ 0.15 $M_i$ is heated to type 4 – 5, and the remainder is type 3 or unheated. 1.5 $M_i$ of the parent body ($1.5\times10^{-5}$ $M_{pb}$) is ejected from the parent body. The mean cooling rate for both impactor and parent body is faster than for $\phi = 50\%$: $\mu_i = 6.56^{+9.84}_{-3.93}$ K/Ma and $\mu_{pb} = 5.32^{+5.31}_{-2.66}$ K/Ma. These cooling rates are greater than for $\phi = 0.5$ because of the increase of thermal conductivity with decreasing $\phi$.

For $\phi = 0$, no material is heated to type 7. Of the impactor, < 0.02 $M_i$ is heated to type 6, and ~ 0.45 $M_i$ is heated to type 4 – 5. 0.28 $M_i$ is heated to type 3, and 0.26 $M_i$ is ejected during the collision. The mass of the material heated from the parent body is much less than for $\phi = 0.2$ or 0.5, with < 0.04 $M_i$ of material heated to type 4 – 5. The low level of heating in this (no porosity) case is consistent with the results of the Keil et al. (1997) who found minimal amounts of melt in their simulations. More of the parent body is ejected in this collision than in the porous collisions: a mass of 3.5 $M_i$ ($3.5\times10^{-5}$ $M_{pb}$) is ejected. The average cooling rate is much higher than for $\phi = 0.2$ or 0.5. The median cooling rate is 16.05 K/Ma for the impactor and 17.25 K/Ma for the parent body. However, some material is cooled even more rapidly: The mean rate for the impactor, $\mu_i$, is $30.01^{+89.32}_{-22.47}$ K/Ma, and for the parent body, $\mu_{pb} = 36.46^{+164.67}_{-29.85}$ K/Ma. This increase in cooling rate is due to two factors: first, the burial depth is typically less for the non-porous case, with the heated impactor spread out more radially, closer to the surface (compare Figure 2(b) with Figure 2(a) and Figure 1(e)); second, the material that is buried is covered by fully compacted material. The lack of pore space in the insulating layer increases the thermal conductivity, and hence heat can escape from the surface much more easily. Less of the impactor mass is retained on the surface ($M_{ret} = 0.74$ $M_i$) compared to the porous collisions: for $\phi = 0.2$, $M_{ret} = 0.93$ $M_i$ and for $\phi = 0.5$, $M_{ret} = 1.00$ $M_i$.

Thus, porosity has several important effects on the thermal evolution of an impacted body. The first two effects are on the dynamics of the collision: High porosity planetesimals prevent much of the heated impactor and parent body material being ejected from the parent body during the collision. The porosity also attenuates the shock wave, resulting in more of the impact energy being concentrated in a region proximal to the impact site. This leads to a third effect, that higher temperatures are reached in highly porous materials than low porosity, as discussed in Davison et al. (2010). The final effect is that the porous rock around the heated material serves to insulate the heated material, slowing its cooling rate and thus keeping materials hot for a longer period of time than in the fully consolidated material that had been used in impact simulations to date.

*3.3.2. $R_i$ = 2.5 km and $v_i$ = 4 km/s*

Figure 7 and Tables 1 and 2 show the peak temperatures and cooling rates for the three simulations with $R_i$ = 2.5 km and $v_i$ = 4 km/s.

Peak temperatures for $R_i$ = 2.5 km are similar to those for $R_i$ = 25 km, with the exception that in the non-porous case, no material is shock heated to type 6 for the smaller impactor. However, the cooling rates for the smaller impactor case are several orders of magnitude higher than for the larger impactor. For example, for $\phi$ = 0.5, the cooling rate in the impactor is $\mu_i$ = $145.22^{+155.93}_{-75.19}$ K/Ma and in the parent body, $\mu_{pb}$ = $167.45^{+245.35}_{-99.52}$ K/Ma, and in the 20% porosity case, $\mu_i$ = $661.23^{+929.36}_{-386.35}$ K/Ma, and $\mu_{pb}$ = $783.19^{+957.19}_{-430.75}$ K/Ma. This reflects the fact that the heated material is much closer to the surface of the parent body, and the buried impactor has a thinner insulating layer covering it in the collision with $R_i$ = 2.5 km. In the non-porous collision, very little heated material is buried, and the lack of pore space means that heat is easily conducted to the surface and radiated away: the cooling rates in the impactor are $\mu_i$ = $2.0 \times 10^5$ $^{+1.4 \times 10^6}_{-1.7 \times 10^5}$ K/Ma, and in the parent body $\mu_{pb}$ = $4.0 \times 10^5$ $^{+3.2 \times 10^6}_{-3.6 \times 10^5}$ K/Ma. These results suggest that the very fast cooling H chondrites (with cooling rates of > 1000 K/Ma; e.g. Taylor et al., 1987) may have been heated by impact and emplaced near to the surface of the H chondrite parent body.

The trend for less of the impactor to be retained and less of the parent body to be ejected at low porosity is the same for both impactor sizes: In the collision with $\phi$ = 0.5, all of the impactor is retained on the parent body, and 2.4 impactor masses of parent body material are ejected ($2.4 \times 10^{-8}$ $M_{pb}$). For $\phi$ = 0.2, 0.07 $M_i$ of the impactor and 4.8 $M_i$ ($4.8 \times 10^{-8}$ $M_{pb}$) of the parent body is ejected, and for $\phi$ = 0, 0.32 $M_i$ of the impactor and 11.2 $M_i$ ($1.1 \times 10^{-7}$ $M_{pb}$) of the parent body is ejected.

*3.3.3. $R_i$ = 25 km and $v_i$ = 6 km/s*

In Figure 8 and Tables 1 and 2, the cooling rates and peak temperatures of the three simulations with $R_i$ = 25 km and $v_i$ = 6 km/s are presented.

Peak temperatures after a collision at 6 km/s are much higher than in the equivalent collision at 4 km/s. For example, in the $\phi$ = 0.5 simulation, 0.75 $M_i$ of the impactor is heated above the dunite liquidus (labeled 'melt' on Figure 8). In both the $\phi$ = 0.2 and $\phi$ = 0 simulations, some material is heated above the solidus; in the equivalent simulations at $v_i$ = 4 km/s, peak temperatures are type 7 ($\phi$ = 0.2) and type 6 ($\phi$ = 0).

Typically, more burial of heated material took place in the faster velocity collisions. This extra burial depth means that the average cooling rates are slower than in the 4 km/s collisions. For example, for $\phi = 0.5$, $\mu_i = 0.79^{+0.52}_{-0.32}$ K/Ma and $\mu_{pb} = 0.54^{+0.27}_{-0.18}$ K/Ma, and for $\phi = 0.2$ $\mu_i = 6.48^{+19.97}_{-4.89}$ K/Ma and $\mu_{pb} = 3.13^{+2.47}_{-1.38}$ K/Ma. For $\phi = 0$, cooling rates have a wide range of values for both the impactor and parent body. Two populations of cooling rates exist, with some material cooling rapidly (> 1000 K/Ma) and the majority cooling at 10 – 100 K/Ma. This divide in the cooling rates is likely due to some heated material being exposed at the surface, and hence losing heat much more quickly than the buried, heated material. This is reflected in the large standard deviation in the cooling rates: $\mu_i = 121.26^{+817.55}_{-105.60}$ K/Ma and $\mu_{pb} = 70.25^{+472.94}_{-61.16}$ K/Ma.

*3.3.4. $R_i = 2.5$ km and $v_i = 6$ km/s*

The final suite of simulations modeled collisions with $R_i = 2.5$ km and $v_i = 6$ km/s, for each initial planetesimal porosity ($\phi = 0$, 0.2, 0.5). Cooling rates and peak temperatures are presented in Figure 9 and Tables 1 and 2.

The general trends for slower cooling rates and higher peak temperatures are also seen in these simulations when compared to the equivalent collisions with $v_i = 4$ km/s. In the collision with $\phi = 0.5$, the average cooling rates are $\mu_i = 47.07^{+6.10}_{-5.40}$ K/Ma and $\mu_{pb} = 48.46^{+47.77}_{-24.06}$ K/Ma. These slower cooling rates are a consequence of the deeper burial and emplacement of the heated material.

Much less heating was observed with $\phi = 0$. Heated material is more proximal to the surface than for $\phi = 0.2$ or 0.5 (with $v_i = 6$ km/s), so cooling rates are more rapid than in those collisions: $\mu_i = 2190.37^{+3811.67}_{-1381.02}$ K/Ma and $\mu_{pb} = 1929.89^{+2572.30}_{-1102.63}$ K/Ma. Unlike the equivalent collision at $v_i = 4$ km/s, there is some burial of the heated impactor, so the cooling rates are not as high as in that simulation.

*3.3.5. $v_i = 2$ km/s*

Simulations were also run with $v_i = 2$ km/s. However, at this low velocity, little heating occurred (as shown in Davison et al. 2010): No material reached the peak temperature required to define it as a type 4, and in no simulation was more than ~1×10⁻³ $M_i$ heated to 773 K. Therefore a detailed analysis of the cooling rates and peak temperatures for these simulations was not performed.

# 4. DISCUSSION

The initial porosity of the planetesimal and the relative collision velocity both play a role in determining the peak temperatures and the cooling rates of the impact processed material. Increasing porosity had several complementary effects: First, higher post shock temperatures were possible, due to the impact energy being concentrated in a smaller region because some of it was spent compacting pore space. Second, lower cooling rates were observed because the layer covering the heated material was more thermally insulating, and third, more of the heated impactor material was retained on the parent body rather than ejected during the collision event.

Increasing the impact velocity has three major effects: First, higher velocities lead to more heating, due to the increased impact energy. Second, burial of a larger mass of heated material was possible, due to the formation of a larger impact crater that collapsed more dramatically under gravity, forming a thicker insulating layer. This deeper burial of material led to the third effect: lower cooling rates, due to heat being less able to escape through the thick insulating layer.

The retention of impactor material on the parent body during the collision is important to the heat budget, as the impactor material and a comparable mass of parent body material, being exposed to the greatest shock compression, is the most strongly heated material. Typically, cooling rates in the impactor and parent body were of a similar magnitude as the cooling rates were largely controlled by the depth of the burial of heated material.

In certain cases, impact melt was produced, which is consistent with recent findings that impact melt is widespread in meteoritic material (Bischoff et al. 2006; Weirich et al. 2010). This melting affected the impactor more than the parent body, although small amounts of parent body material were melted. Wasson et al. (1980) and Choi et al (1995) suggested that the IAB iron meteorites may have formed from localized impact melt pools on a parent body. However, Benedix et al. (2000) argued that as impact heating was likely to be inefficient (Keil et al, 1997) another mechanism was required to heat the parent body prior to fragmentation by impact. Our results are compatible with the creation of isolated pools of impact melt, which suggests that it may be possible to form some IAB irons by impact processes. Cooling rates of ~ 1 – 10 K/Ma are required for the IAB irons (Winfield et al., 2012), which is consistent with many of the cooling rates observed in our model. Further study of impacts involving iron is required to fully understand the complex formation of the IAB meteorites.

Of the 12 simulations described above, 10 have mean cooling rates in the range 1 - 1000 K/Ma, which match the observations from Taylor et al. (1987), Willis and Goldstein (1981) and Williams et al. (2000). Of the two simulations that have mean cooling rates that fall outside that range ($R_i$ = 2.5 km, $\phi$ = 0, $v_i$ = 4 km/s and 6 km/s), the simulation with $v_i$ = 6 km/s does have cooling rates within one $\sigma$ of the mean that lie between 1 - 1000 K/Ma. Moreover, the iSALE simulations do not account for the dilatancy of the planetesimal material. Therefore, any porosity increase that would have been caused by shear in the collapsing debris lens is neglected; if porosity was introduced to the debris lens in the non-porous simulations, the insulating effect of this lens would be greater, and cooling rates may be slower. Accounting for dilation in iSALE (e.g. Collins et al. 2011b) will allow this process to be investigated in a future study. We conclude that in many different impact scenarios for a range of impact velocities, porosities and impactor sizes, it is possible for material heated by a single impact event to cool at a rate consistent with measurements from a wide range of meteoritic samples.

A simulation was run with a smaller parent body ($R_{pb}$ = 50 km), with an impactor to parent body radius ratio of 0.1 (i.e. $R_i$ = 5 km), and a velocity, $v_i$ = 4 km/s. Cooling rates were higher than for the equivalent collisions with $R_{pb}$ = 250 km: for $\phi$ = 0, the mean cooling rate was $\mu$ = $2222.35^{+645.03}_{-499.93}$ K/Ma, for $\phi$ = 0.2, $\mu$ = $285.84^{+953.29}_{-219.90}$ K/Ma, and for $\phi$ = 0.5, $\mu$ = $11.42^{+30.15}_{-8.28}$ K/Ma. While these cooling rates represent a significant increase from the cooling rates on the larger parent body, the cooling rates on the small porous parent body still fits within the range seen from chondritic materials. As smaller bodies are likely to support a larger porosity (as their internal pressures would have been lower), this result does not change our conclusions. An investigation into thermal histories on a range of parent body sizes and structures will be discussed in future work.

To test the sensitivity of the model to the thermal parameters used, an additional thermal evolution simulation was run for the case of $R_i$ = 25 km and $v_i$ = 4 km/s with a different value for thermal diffusivity – the value of $\kappa$ = $7 \times 10^{-3}$ cm$^2$/ was selected as an end member of the range of values that have been employed in other studies. This value was derived from the thermal conductivity used by Opeil et al. (2010), the heat capacity from Ghosh et al. (1999), and the density used in our model. The same output from the iSALE simulation was used in this simulation as was used for the simulation with $\kappa$ = $1 \times 10^{-3}$ cm$^2$/s. There was negligible change in the peak temperatures of the material from both the impactor and the parent body. However, the material that was heated above 773 K cooled more rapidly than in the simulation presented above (Table 1). For the impactor, $\mu_i$ = $45.94^{+64.99}_{-26.91}$ K/Ma and for the parent body, $\mu_{pb}$ = $36.98^{+30.52}_{-16.72}$ K/Ma. Both of these means represent an increase of a factor of ~ 7, corresponding to the change in thermal diffusivity. While this increase in cooling rates is not negligible, it is less than an order of magnitude change, and the cooling rates are still within the measured values for H chondrites (Taylor et al., 1987; Willis and Goldstein, 1981; Williams et al., 2000). More work is required to more precisely constrain the thermal diffusivity appropriate for Solar System materials.

While Keil et al (1997) and Davison et al. (2010) rule out global heating by impacts (i.e. significantly increasing the globally averaged temperature of the parent body) as a heat source, the simulations presented here show that individual impact events could produce materials that match both the types of peak temperatures recorded in meteoritical samples, and the cooling rates in those samples. As most meteorite samples that have been studied are on the centimeter to meter scale, there is no requirement for a heat source to globally heat a parent body and produce the peak temperatures and cooling rates seen in the sample. Moreover, recent N-body simulations (Bottke et al. 2005; O'Brien et al. 2006), semi-analytic modeling (Chambers 2006) and Monte Carlo simulations (Davison et al. 2011) suggest that multiple impacts on a parent body would have been likely in the first 100 Ma. Several studies of the onion shell model have concluded that to fit all measurements of peak temperatures, cooling rates and closure times, a process which perturbs the structure of the parent body is required (e.g. Harrison and Grimm 2010; Taylor et al. 1987; Scott et al. 2010; 2011), with the most likely candidate being impacts. Multiple non-disruptive impacts on the surface of a parent body could not only disturb the onion shell model in the outermost layers of the parent body but also provide an extra heat source to that processed material. This could reduce the depth that those studies required material to have been excavated from, as the impact could increase the petrologic type of the locally disturbed material.

Rubin (2004) studied 210 ordinary chondrites with petrologic types 4 – 6 and shock stages S1 and S2, and concluded that all equilibrated OCs were shocked to shock stages S3 – S6. Many ordinary chondrites retained that high shocked state (and received no further thermal processing). For those that are observed today with shock stages lower than S3, it was proposed that post-shock annealing processes returned them to S1 (with further shock events required to reach S2). It was speculated in that study that burial of impact heated porous rocks could have led to such annealing processes. Other groups of meteorites also exhibit evidence of being shock heated and annealed (e.g. Rubin, 1992; Rubin 2006; Rubin and Wasson 2011). The impact scenarios presented in this work show quantitatively that collisions could both heat and bury material such that it would remain hot for millions of years after the impact event. Whether annealing actually occurred as required by the scenario outlined by Rubin (2004) requires further study. However, it is clear that the thermal energy of the impact can be retained for timescales of millions to hundreds of millions of years, allowing for the possibility of significant parent body processing. Given the temperatures and timescales, it is possible that impacts could have led to the mobilization of volatiles on the parent body or aqueous activity through the localized melting of ice, resulting in a pool of warm water that could react with the rock in the planetesimal. These issues will be investigated in future work.

Yang et al. (2011) presented several effects of shock on iron meteorites, for example the loss of the cloudy zone microstructure, grain boundary diffusion, and localized diffusion of Ni at the kamacite/taenite boundary; all these microscopic effects most likely took place by low temperature shock heating, similar to that discussed in this paper. However, as the results in this work exclusively used dunite as a material, further modeling of impacts on bodies containing iron are required to extend these results to apply to the meteorites studied by Yang et al. (2011), and test their hypothesis.

# 5. CONCLUSIONS

We have modeled the process of impact heating on a spatial scale of 0.1 to 1 km and the subsequent post-impact thermal evolution of a porous body struck by an impactor a relatively low velocities. These simulations show that impacts can heat chondritic material to temperatures that would produce a range of petrologic types; from type 3 through to impact melt. The heated material subsequently cooled at rates consistent with measurements made from meteoritic samples (i.e. ~ 1 – 1000 K/Ma at 773 K). A thorough examination of impact heating coupled with heating from radiogenic decay of $^{26}$Al is now required, to fully quantify the interplay between those two heat sources and the effects of the thermal evolution of the chondritic parent bodies.

*Acknowledgments:* We gratefully acknowledge the major contributions of Jay Melosh, Boris Ivanov, Kai Wünnemann and Dirk Elbeshausen to the development of iSALE, and Laurence Billingham for his contribution to the self-gravity algorithm. We thank Gregory Herzog, Ed Scott, Alan Rubin and an anonymous reviewer, whose comments improved this paper. TMD and FJC were funded by NASA PGG grant NNX09AG13G. GSC was supported by NERC grant NE/E013589/1 and STFC grant ST/G002452/1.

Table 1: Median and mean cooling rates (at 773 K) for each impact simulation. In all cases presented here, the parent body had an initial radius of 250 km. Median cooling rates were calculated as a mass-weighted median for all cells that are heated above 773 K. Mean cooling rates were calculated as the geometric mean. One simulation was run with a different thermal diffusivity, to test the sensitivity to model parameter. In that case, $\kappa = 7\times10^{-3}$ cm$^2$/s was used (derived from Opeil et al., 2010; Ghosh et al., 1999). In all other cases, $\kappa = 1\times10^{-3}$ cm$^2$/s was used (Ghosh et al. 2003).

| Initial conditions | | | | Impactor | | | | Parent body | | | |
|---|---|---|---|---|---|---|---|---|---|---|---|
| $R_i$ [km] | $\phi$ | $v_i$ [km/s] | $\kappa \times 10^3$ [cm$^2$/s] | Cooling rate [K/Ma] | | | | Cooling rate [K/Ma] | | | |
| | | | | Median | $\mu_i$ | $+\sigma_i$ | $-\sigma_i$ | Median | $\mu_{pb}$ | $+\sigma_{pb}$ | $-\sigma_{pb}$ |
| 25.0 | 0.0 | 4 | 1 | 16.05 | 30.01 | 89.32 | 22.47 | 17.25 | 36.46 | 164.67 | 29.85 |
| 25.0 | 0.2 | 4 | 1 | 4.90 | 6.56 | 9.84 | 3.93 | 4.62 | 5.32 | 5.31 | 2.66 |
| 25.0 | 0.2 | 4 | 7 | 34.28 | 49.94 | 64.99 | 26.91 | 32.36 | 36.98 | 30.52 | 16.72 |
| 25.0 | 0.5 | 4 | 1 | 1.08 | 1.16 | 0.27 | 0.22 | 1.00 | 1.00 | 0.46 | 0.32 |
| 25.0 | 0.0 | 6 | 1 | 79.55 | 121.26 | 817.55 | 105.60 | 27.58 | 70.25 | 472.94 | 61.16 |
| 25.0 | 0.2 | 6 | 1 | 3.82 | 6.48 | 19.97 | 4.89 | 2.85 | 3.13 | 2.47 | 1.38 |
| 25.0 | 0.5 | 6 | 1 | 0.67 | 0.79 | 0.52 | 0.32 | 0.52 | 0.54 | 0.27 | 0.18 |
| 2.5 | 0.0 | 4 | 1 | $8.8\times10^4$ | $2.0\times10^5$ | $1.4\times10^6$ | $1.7\times10^5$ | $3.1\times10^5$ | $4.0\times10^5$ | $3.2\times10^6$ | $3.6\times10^5$ |
| 2.5 | 0.2 | 4 | 1 | 498.40 | 661.23 | 929.36 | 386.35 | 596.40 | 783.19 | 957.19 | 430.75 |
| 2.5 | 0.5 | 4 | 1 | 130.40 | 145.22 | 155.93 | 75.19 | 128.80 | 167.45 | 245.35 | 99.52 |
| 2.5 | 0.0 | 6 | 1 | 1352.00 | 2190.37 | 3811.67 | 1391.02 | 1348.40 | 1929.89 | 2572.30 | 1102.63 |
| 2.5 | 0.2 | 6 | 1 | 308.40 | 349.22 | 228.83 | 138.24 | 324.40 | 469.82 | 476.83 | 236.65 |
| 2.5 | 0.5 | 6 | 1 | 47.60 | 47.07 | 6.10 | 5.40 | 42.80 | 48.46 | 47.77 | 24.06 |

Table 2: Mass of material originating from the impactor and parent body in each simulation, that has a peak temperature in 6 different petrologic type temperature ranges. The peak temperature of each computational cell is used to define the petrologic type of that mass of material, using the definition of Harrison & Grimm (2010) for type 3 (<948 K), type 4-5 (948-1138 K) and type 6 (1138-1273 K). We define type 7 as mass heated above 1273 K, but below the dunite solidus (1373 K; McKenzie and Bickle, 1988), and we use 2053 K as the dunite liquidus (Katz et al., 2003). Heated mass fractions quoted in this table are defined as: for the impactor, the fraction of the impactor mass heated to a given petrologic type for at least one year, and retained on the parent body; for the parent body, the mass of material heated to the given petrologic type for at least one year and retained on the parent body, normalized by the initial mass of the impactor, $M_i$.

| $R_i$ [km] | $\phi$ | $v_i$ [km/s] | Body | Petrologic Type: 3 / Temp 293-948 | 4-5 / 948-1138 | 6 / 1138-1273 | 7 / 1273-1373 | Solid + Melt / 1373-2053 | Complete melting / 2053+ |
|---|---|---|---|---|---|---|---|---|---|
| 25 | 0.0 | 4 | Impactor | $2.8\times10^{-1}$ | $4.5\times10^{-1}$ | $1.6\times10^{-2}$ | 0.0 | 0.0 | 0.0 |
| 25 | 0.0 | 4 | Parent body | $9.9\times10^{2}$ | $3.9\times10^{-2}$ | $1.4\times10^{-3}$ | 0.0 | 0.0 | 0.0 |
| 25 | 0.2 | 4 | Impactor | $6.8\times10^{-2}$ | $9.2\times10^{-2}$ | $7.6\times10^{-1}$ | $1.3\times10^{-2}$ | 0.0 | 0.0 |
| 25 | 0.2 | 4 | Parent body | $9.9\times10^{2}$ | $1.5\times10^{-1}$ | $4.1\times10^{-3}$ | $7.3\times10^{-5}$ | 0.0 | 0.0 |
| 25 | 0.5 | 4 | Impactor | $2.2\times10^{-2}$ | $8.2\times10^{-2}$ | $6.4\times10^{-2}$ | $7.2\times10^{-1}$ | $1.2\times10^{-1}$ | 0.0 |
| 25 | 0.5 | 4 | Parent body | $9.9\times10^{2}$ | $3.6\times10^{-1}$ | $2.5\times10^{-2}$ | $7.9\times10^{-3}$ | $8.2\times10^{-4}$ | 0.0 |
| 25 | 0.0 | 6 | Impactor | $2.1\times10^{-1}$ | $5.3\times10^{-2}$ | $6.8\times10^{-2}$ | $2.2\times10^{-1}$ | $1.0\times10^{-1}$ | 0.0 |
| 25 | 0.0 | 6 | Parent body | $9.9\times10^{2}$ | $8.8\times10^{-2}$ | $1.9\times10^{-2}$ | $5.2\times10^{-3}$ | $4.8\times10^{-2}$ | 0.0 |
| 25 | 0.2 | 6 | Impactor | $9.1\times10^{-2}$ | $2.7\times10^{-2}$ | $4.6\times10^{-2}$ | $3.2\times10^{-2}$ | $6.7\times10^{-1}$ | 0.0 |
| 25 | 0.2 | 6 | Parent body | $9.9\times10^{2}$ | $5.1\times10^{-1}$ | $1.7\times10^{-1}$ | $4.5\times10^{-2}$ | $1.9\times10^{-2}$ | 0.0 |
| 25 | 0.5 | 6 | Impactor | $1.7\times10^{-2}$ | $1.3\times10^{-2}$ | $2.0\times10^{-2}$ | $2.1\times10^{-2}$ | $1.8\times10^{-1}$ | $7.6\times10^{-1}$ |
| 25 | 0.5 | 6 | Parent body | $9.9\times10^{2}$ | $1.2\times10^{0}$ | $3.8\times10^{-1}$ | $1.6\times10^{-1}$ | $3.1\times10^{-1}$ | $6.2\times10^{-3}$ |
| 2.5 | 0.0 | 4 | Impactor | $4.7\times10^{-1}$ | $2.1\times10^{-1}$ | 0.0 | 0.0 | 0.0 | 0.0 |
| 2.5 | 0.0 | 4 | Parent body | $1.0\times10^{6}$ | $2.1\times10^{-2}$ | 0.0 | 0.0 | 0.0 | 0.0 |
| 2.5 | 0.2 | 4 | Impactor | $8.2\times10^{-2}$ | $9.9\times10^{-2}$ | $7.3\times10^{-1}$ | $2.3\times10^{-2}$ | 0.0 | 0.0 |
| 2.5 | 0.2 | 4 | Parent body | $1.0\times10^{6}$ | $1.1\times10^{-1}$ | $8.2\times10^{-3}$ | $1.7\times10^{-4}$ | 0.0 | 0.0 |
| 2.5 | 0.5 | 4 | Impactor | $4.2\times10^{-2}$ | $8.9\times10^{-2}$ | $7.9\times10^{-2}$ | $7.3\times10^{-1}$ | $6.0\times10^{-2}$ | 0.0 |
| 2.5 | 0.5 | 4 | Parent body | $1.0\times10^{6}$ | $3.1\times10^{-1}$ | $2.2\times10^{-2}$ | $4.7\times10^{-3}$ | $9.3\times10^{-5}$ | 0.0 |
| 2.5 | 0.0 | 6 | Impactor | $1.7\times10^{-1}$ | $5.2\times10^{-2}$ | $3.8\times10^{-2}$ | $1.7\times10^{-1}$ | $9.6\times10^{-2}$ | 0.0 |
| 2.5 | 0.0 | 6 | Parent body | $1.0\times10^{6}$ | $7.3\times10^{-2}$ | $1.0\times10^{-2}$ | $3.0\times10^{-3}$ | $1.6\times10^{-5}$ | 0.0 |
| 2.5 | 0.2 | 6 | Impactor | $5.7\times10^{-2}$ | $3.1\times10^{-2}$ | $5.7\times10^{-2}$ | $5.0\times10^{-2}$ | $6.9\times10^{-1}$ | 0.0 |
| 2.5 | 0.2 | 6 | Parent body | $1.0\times10^{6}$ | $5.7\times10^{-1}$ | $1.2\times10^{-1}$ | $6.1\times10^{-2}$ | $3.8\times10^{-2}$ | 0.0 |
| 2.5 | 0.5 | 6 | Impactor | $4.8\times10^{-3}$ | $2.2\times10^{-2}$ | $1.2\times10^{-2}$ | $1.4\times10^{-2}$ | $2.2\times10^{-1}$ | $7.2\times10^{-1}$ |
| 2.5 | 0.5 | 6 | Parent body | $1.0\times10^{6}$ | $1.5\times10^{0}$ | $3.0\times10^{-1}$ | $1.2\times10^{-1}$ | $2.9\times10^{-1}$ | $4.7\times10^{-3}$ |

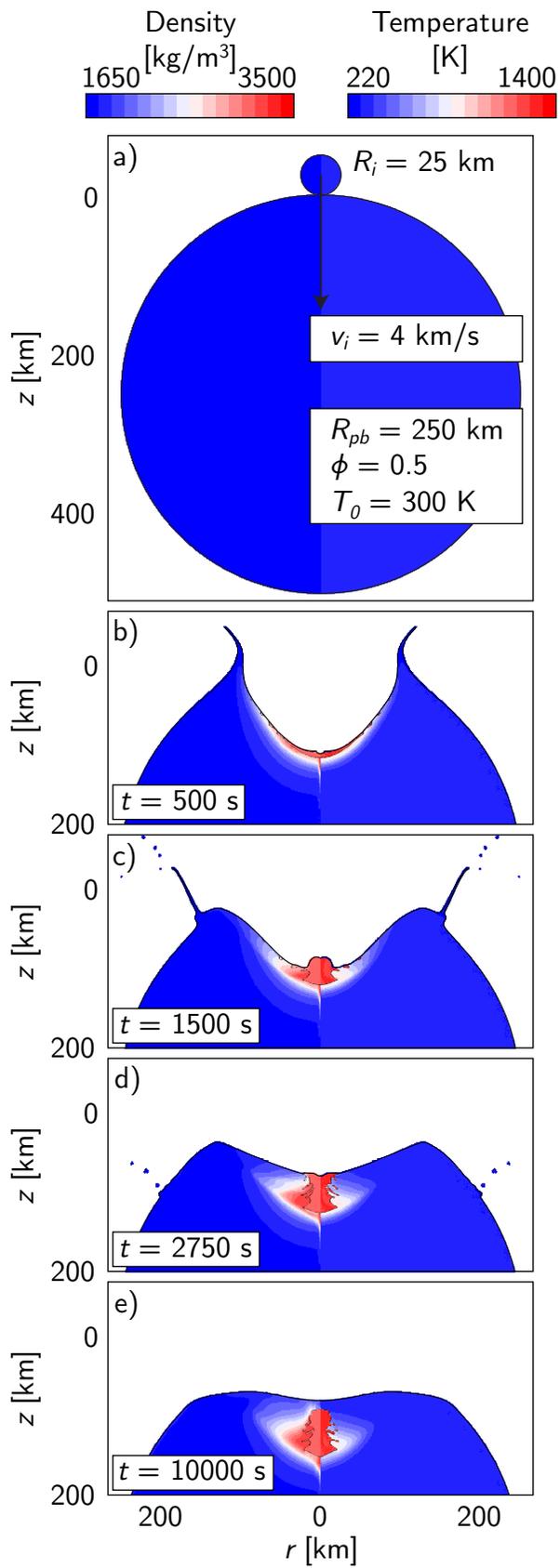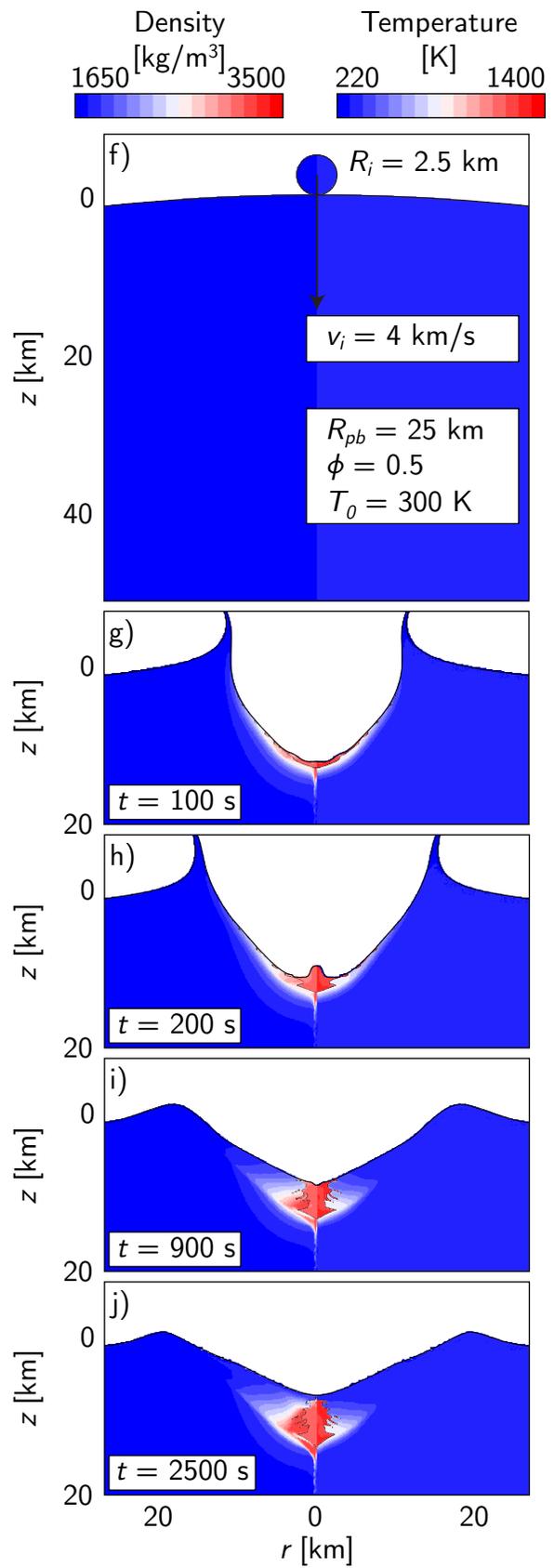

Figure 1: Snapshot at several times during the impact event. Shown here are impacts into a 250 km radius parent body of a 25 km radius impactor (left) and a 2.5 km radius impact (right), with an impact velocity of 4 km/s and an initial porosity of 0.5. Each frame shows a slice through the center of the colliding planetesimals: the left hand side of the frame shows the density structure and the right hand side shows the temperature contours at that time. The final morphology of the crater and thermal anomaly are shown in the final frame. $z = 0$ is the point of impact. A more detailed view of the heated region from Figure 1(e) can be seen in Figure 3(a). A color version of this figure is available in the online version.

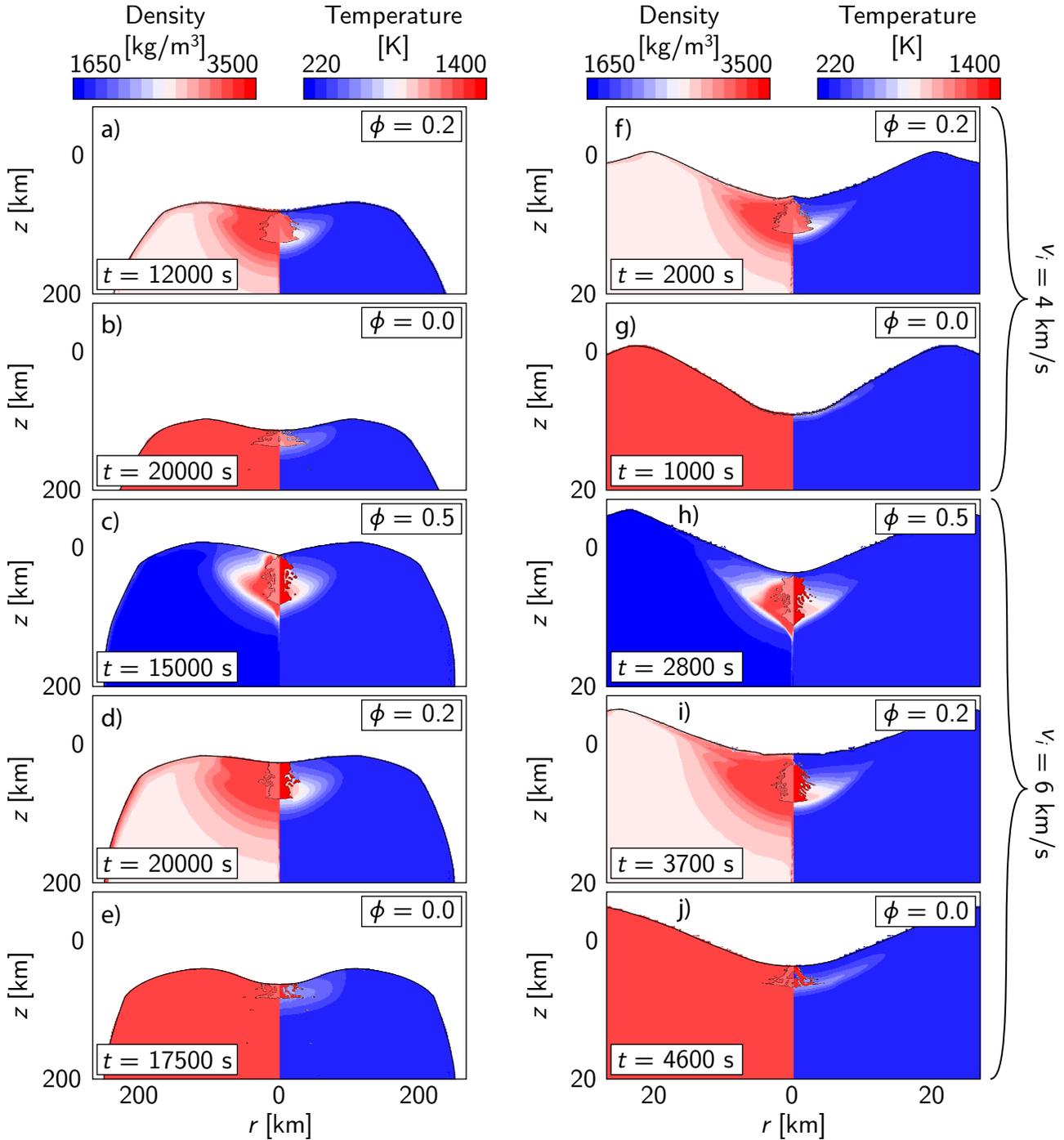

Figure 2: Final crater morphologies for each of the models discussed in this paper, except for those presented in Figure 1. The five frames on the left hand side show craters produced by 25 km radius impactors; the five frames on the right hand side show craters produces by 2.5 km radius impactors. Each frame shows a slice through the center of the crater: the left hand side of the frame shows the density structure and the right hand side shows the temperature contours at that time. $z = 0$ is the point of impact. A color version of this figure is available in the online version.

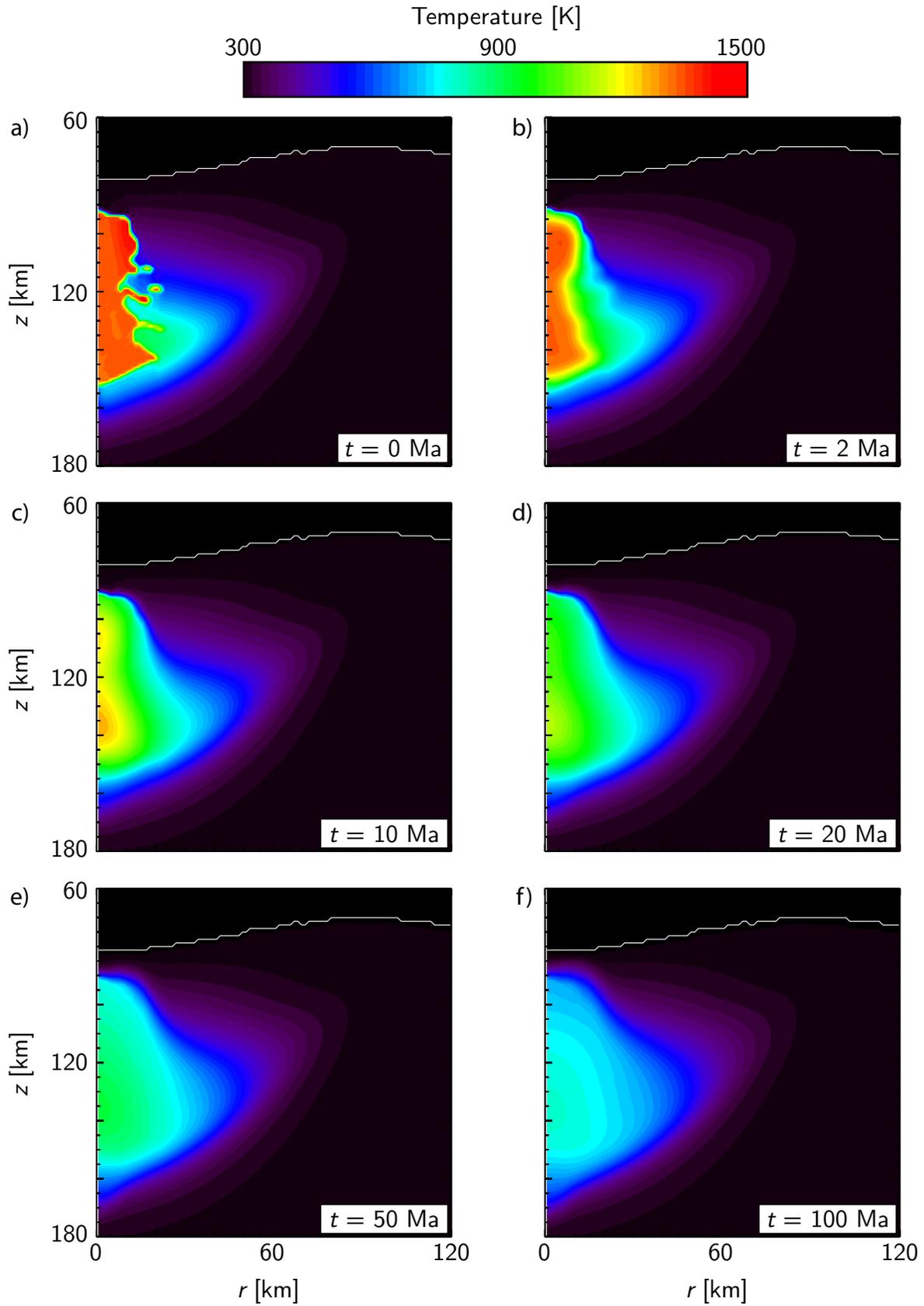

Figure 3: Temperature contours through time of the heated region after the impact presented on the left hand side of Figure 1 – i.e. for a 25 km radius impactor colliding with a 250 km radius parent body at 4 km/s, with an initial porosity of 0.5. Times stated are the time after the impact event.

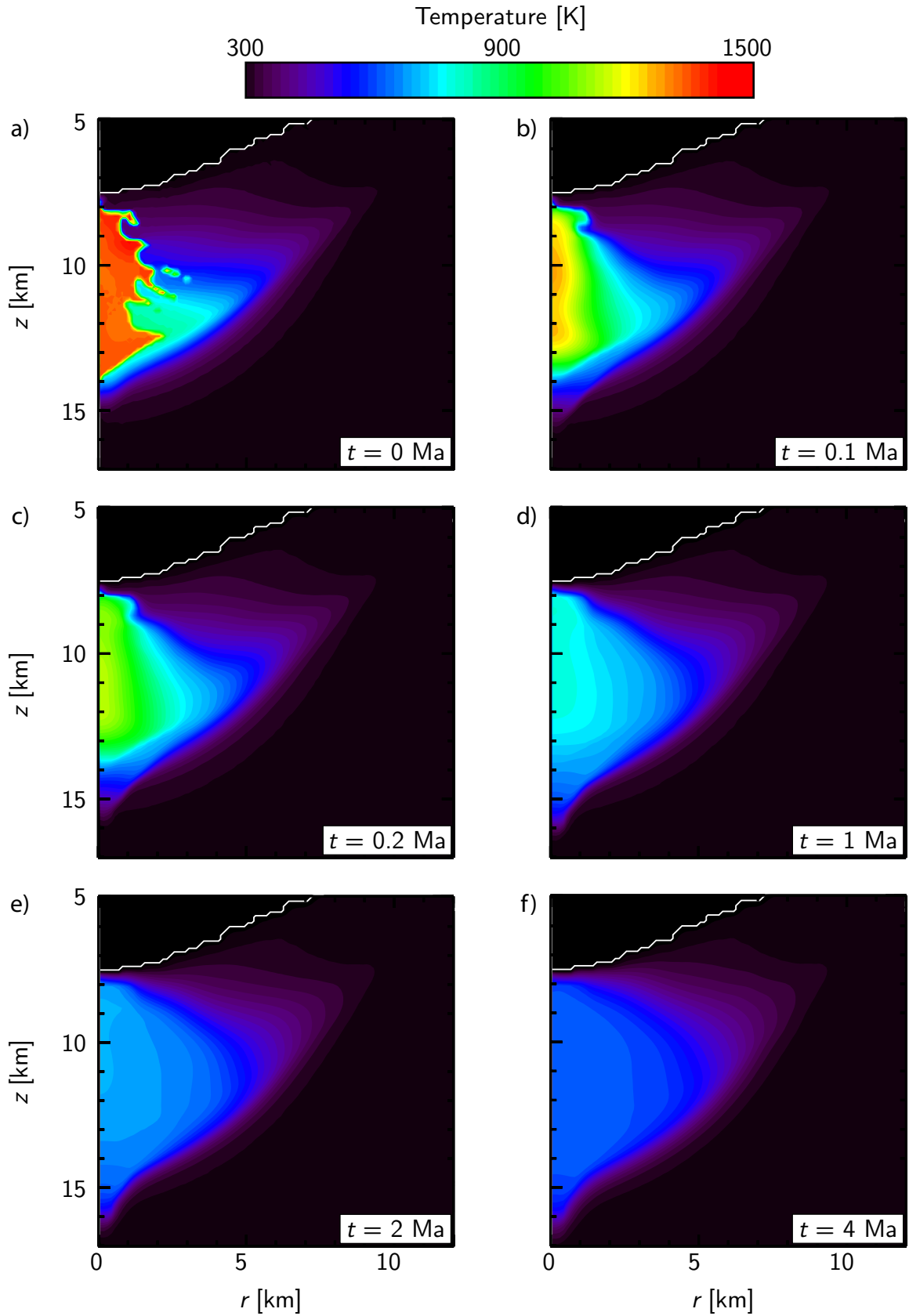

Figure 4: Temperature contours through time of the heated region after the impact presented on the right hand side of Figure 1 – i.e. for a 2.5 km radius impactor colliding with a 250 km radius parent body at 4 km/s, with an initial porosity of 0.5. Times stated are the time after the impact event.

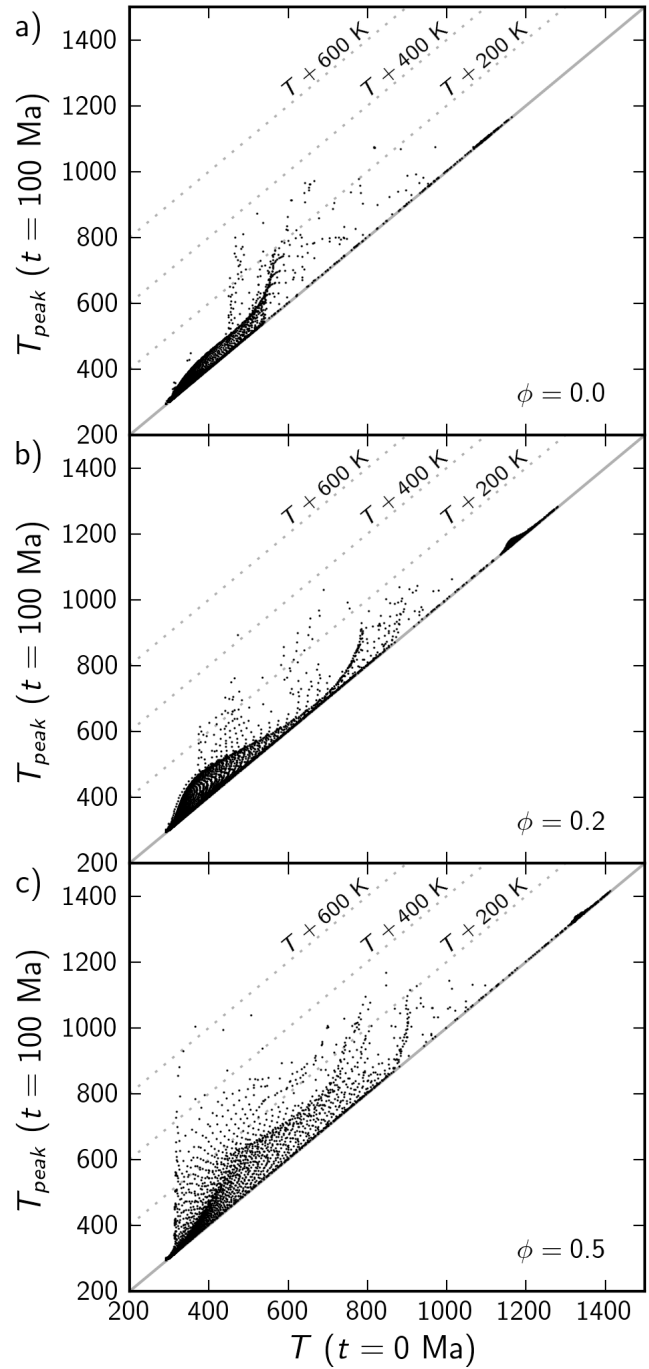

Figure 5: A comparison of the temperature immediately after the impact event, $T(t = 0)$ and the peak temperature reached after 100 Ma of thermal evolution, $T_{peak}(t = 100$ Ma$)$, for three models with $v_i = 4$ km/s and $R_i = 25$ km. Each point represents one computational grid cell. Due to the structure of the axisymmetric mesh, each cell does not represent an equal volume. The dotted lines serve as a guide, and represent an increase of 200, 400 and 600 K from the post impact temperature to the peak temperature. Qualitatively, as porosity increases, there is an increase in the number of cells heated post-impact, and in the level of heating achieved in those cells.

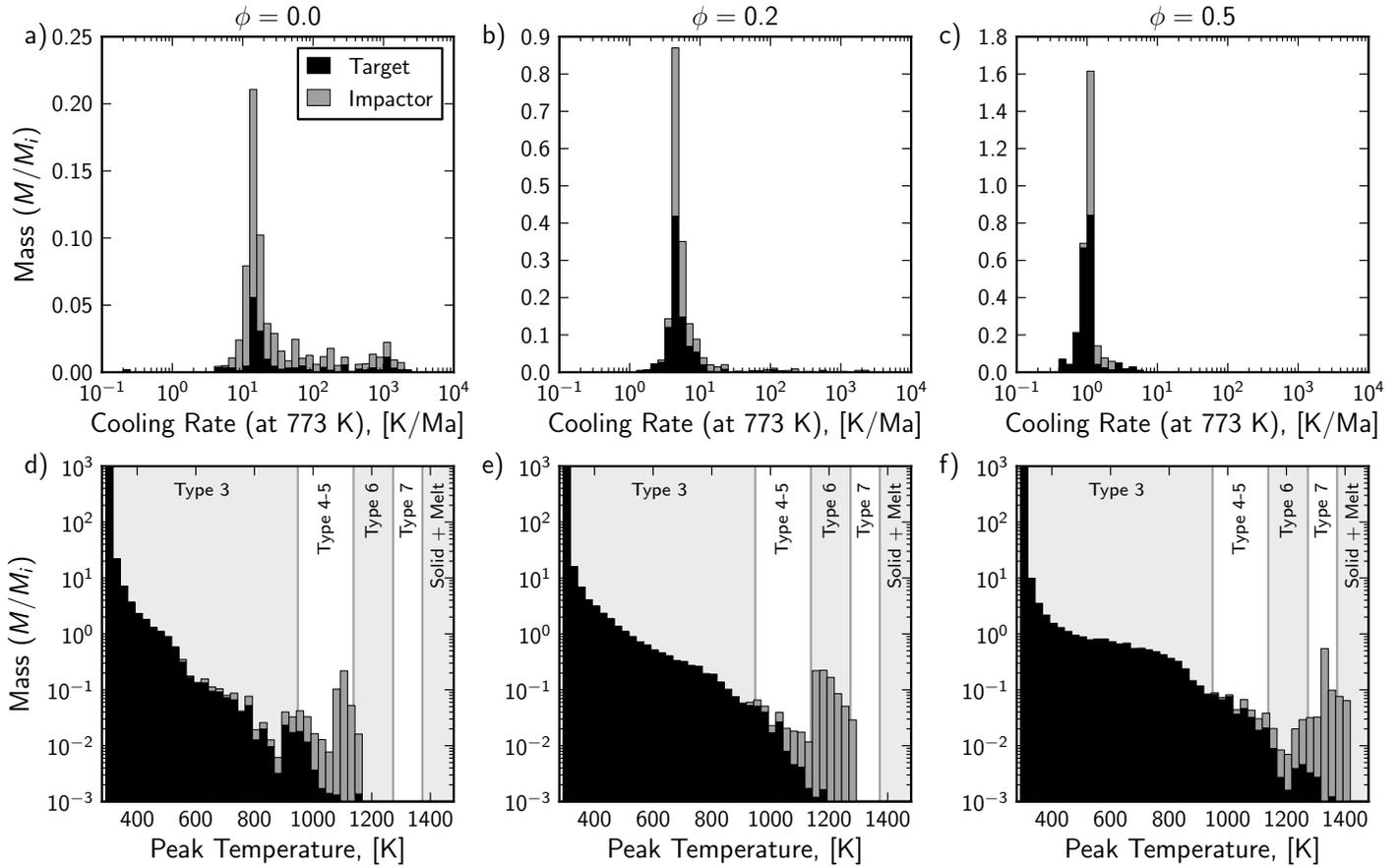

Figure 6: Cooling rates and peak temperatures (after 100 Ma) for collisions between a 25 km radius planetesimal and a 250 km radius planetesimal at 4 km/s, for three porosities ($\phi$ = 0, 0.2, 0.5). Black bars show the mass from the parent body, and grey bars show the mass from the smaller impacting planetesimal (bars are stacked). Shading behind the peak temperature plots shows which petrologic type the material falls into, based on peak temperature estimates from Harrison & Grimm (2010). Note the differing ordinate scales for the different porosities on the cooling rate plots (a-c). For the cooling rate plots (a-c), bins are logarithmic, with a width of 0.1 in $\log_{10}$; for the peak temperature plots (d-f), the bin width is 25 K.

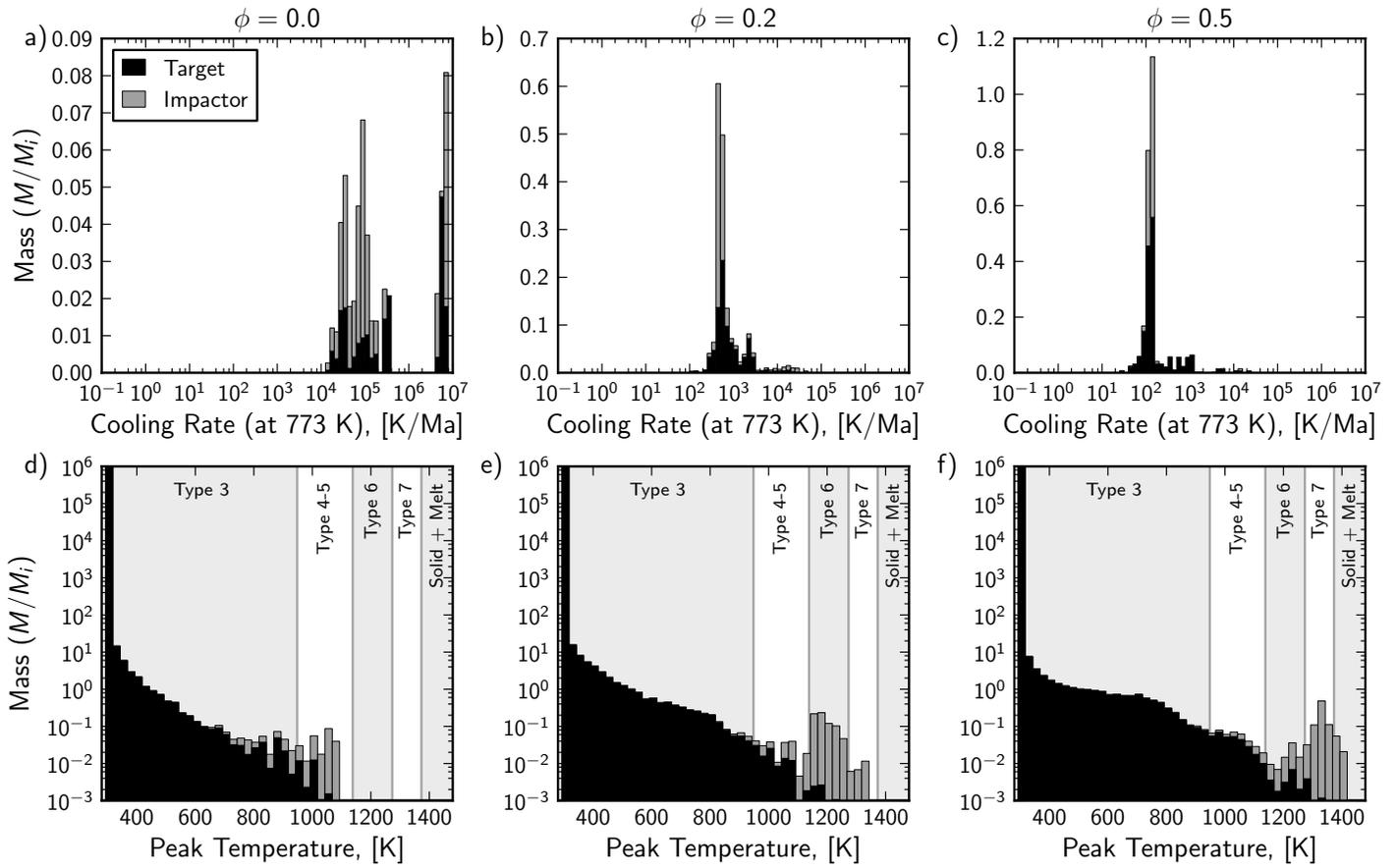

Figure 7: As in Figure 6, but for collisions at 4 km/s between a 2.5 km radius impactor and a 250 km radius parent body. Peak temperatures are for the first 4 Ma after impact.

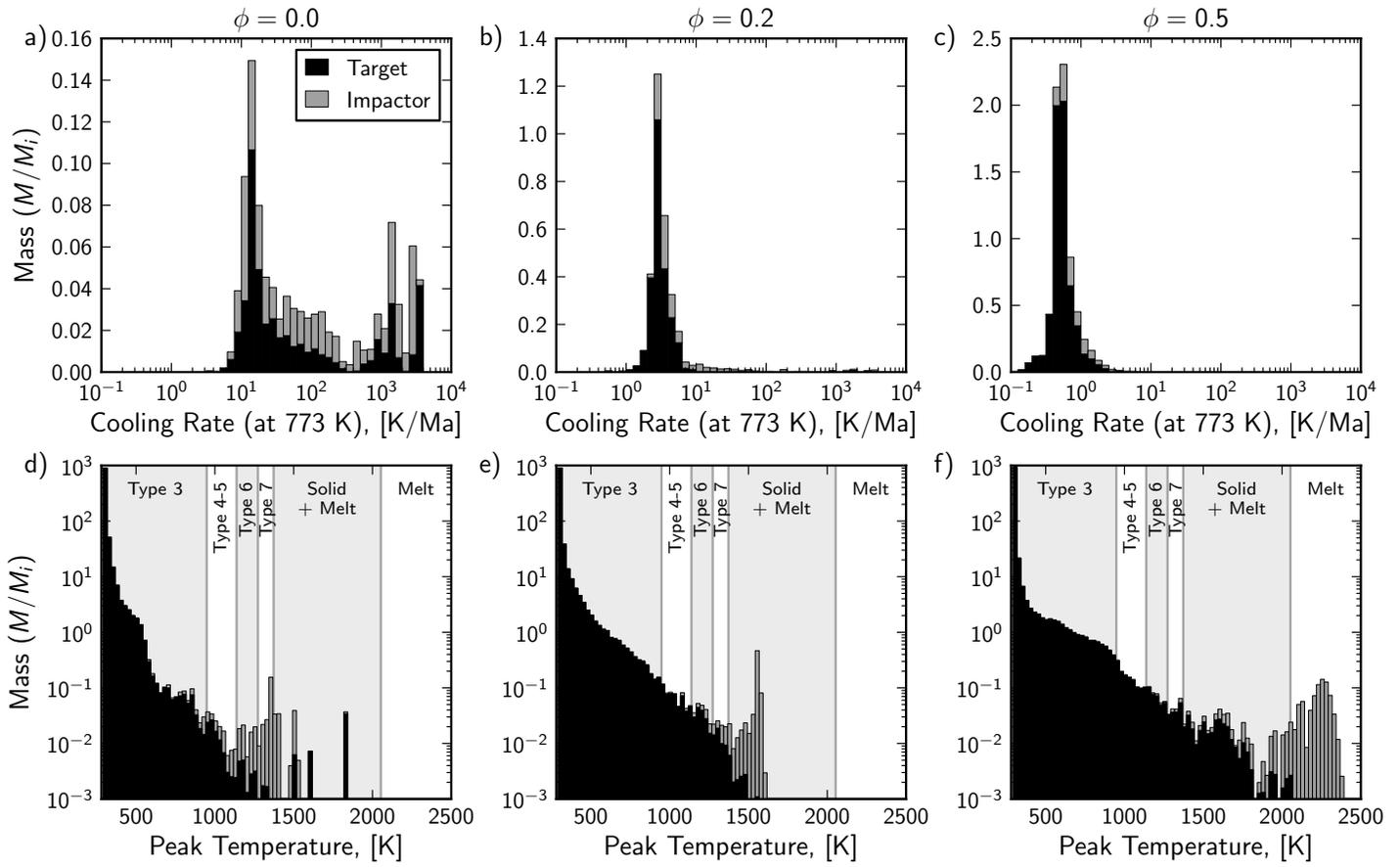

Figure 8: As in Figure 6, but for collisions at 6 km/s between a 25 km radius impactor and a 250 km radius parent body. Peak temperatures are for the first 100 Ma after impact.

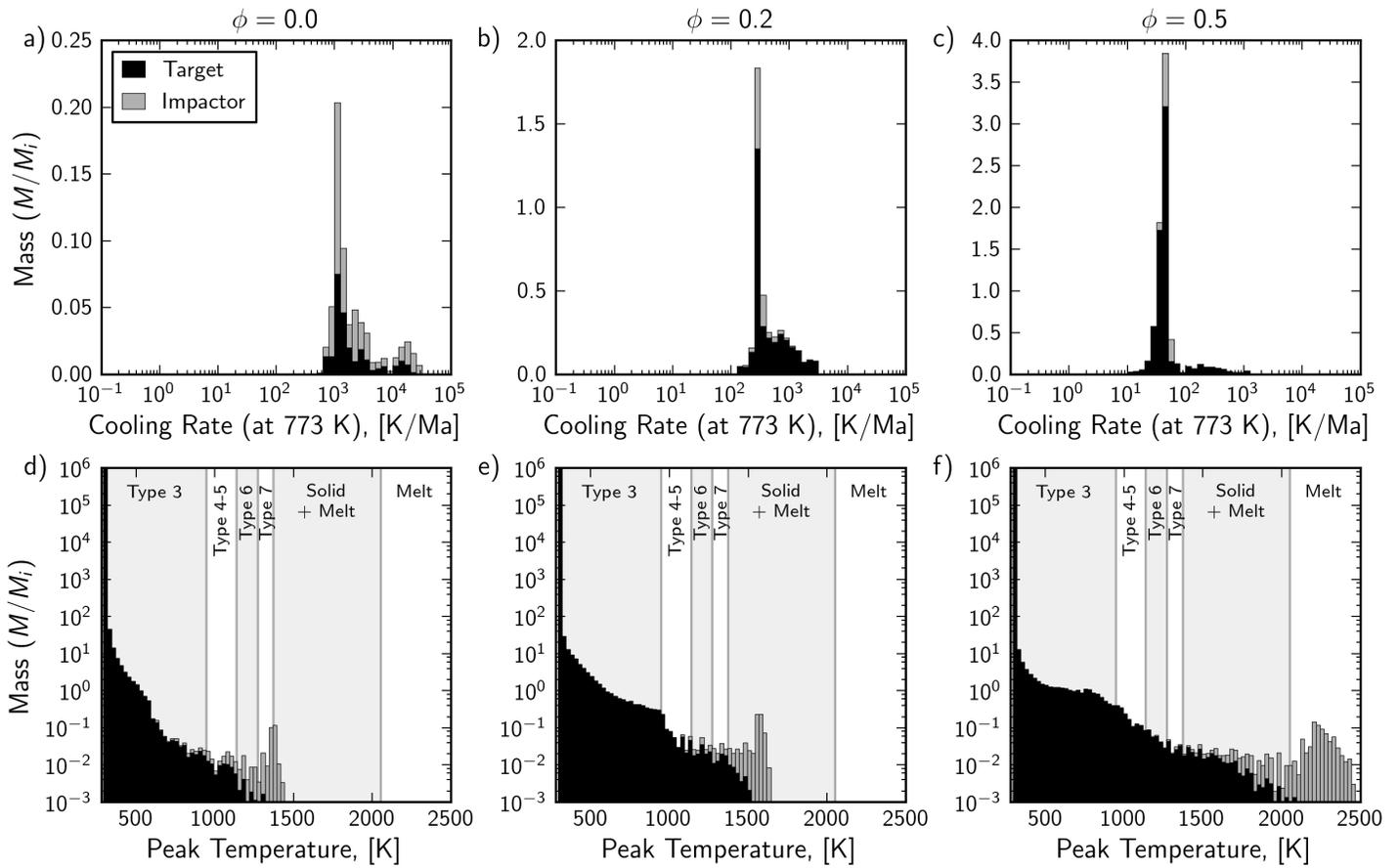

Figure 9: As in Figure 6, but for collisions at 6 km/s between a 2.5 km radius impactor and a 250 km radius parent body. Peak temperatures are for the first 4 Ma after impact.